\documentclass{sigchi}

\toappear{Permission to make digital or hard copies of all or part of this work for personal or classroom use is granted without fee provided that copies are not made or distributed for profit or commercial advantage and that copies bear this notice and the full citation on the first page. \\
 UCL Technical Report. {20 May, 2015, London, UK.}\\
 Copyright \copyright~2015 University College London.}
\usepackage{balance}  % to better equalize the last page
\usepackage{graphics} % for EPS, load graphicx instead
\usepackage{times}    % comment if you want LaTeX's default font
\usepackage{url}      % llt: nicely formatted URLs
\usepackage{subfigure}
\makeatletter
\def\url@leostyle{%
  \@ifundefined{selectfont}{\def\UrlFont{\sf}}{\def\UrlFont{\small\bf\ttfamily}}}
\makeatother
\def\pprw{8.5in}
\def\pprh{11in}

\usepackage[pdftex]{hyperref}
\hypersetup{
pdftitle={SIGCHI Conference Proceedings Format},
pdfauthor={LaTeX},
pdfkeywords={SIGCHI, proceedings, archival format},
bookmarksnumbered,
pdfstartview={FitH},
colorlinks,
citecolor=black,
filecolor=black,
linkcolor=black,
urlcolor=black,
breaklinks=true,
}

\usepackage[margin=0.82in]{geometry}

\begin{document}

\title{An Empirical Study on Display Ad Impression Viewability Measurements}

%\numberofauthors{2}
%\author{
%  \alignauthor 1st Author Name\\
%    \affaddr{Affiliation}\\
%    \affaddr{Address}\\
%    \email{e-mail address}\\
%    \affaddr{Optional phone number}
%  \alignauthor 2nd Author Name\\
%    \affaddr{Affiliation}\\
%    \affaddr{Address}\\
%    \email{e-mail address}\\
%    \affaddr{Optional phone number}
%}

\numberofauthors{1}
\author{
  \alignauthor Weinan Zhang, Ye Pan, Tianxiong Zhou, Jun Wang\\
    \affaddr{Department of Computer Science, University College London}\\
    \email{\{w.zhang, y.pan, j.wang\}@cs.ucl.ac.uk}\\
    \email{txzhow@gmail.com}\\
}

\maketitle

\begin{abstract}
Display advertising normally charges advertisers for every single ad impression. Specifically, if an ad in a webpage has been loaded in the browser, an ad impression is counted. However, due to the position and size of the ad slot, lots of ads are actually not viewed but still measured as impressions and charged. These fraud ad impressions indeed undermine the efficacy of display advertising. A perfect ad impression viewability measurement should match what the user has really viewed with a short memory. In this paper, we conduct extensive investigations on display ad impression viewability measurements on dimensions of ad creative displayed pixel percentage and exposure time to find which measurement provides the most accurate ad impression counting. The empirical results show that the most accurate measurement counts one ad impression if more than 75\% of the ad creative pixels have been exposed for at least 2 continuous seconds.
\end{abstract}

\keywords{
Display Advertising, Impression Measuring, Ad Tracking
}

\category{H.5.m.}{Information Interfaces and Presentation (e.g. HCI)}{Miscellaneous}

\section{Introduction}

Display advertising, i.e. displaying ad creative images on the webpages, is one of the mainstream advertising formats for the current popular Internet advertising. During Q4 2013 in US, the display advertising accounted for \$3.7 billion, i.e. 30\% of total revenues \cite{IAB2013}. Moreover, the recently emerged real-time bidding (RTB) techniques act as a new paradigm for display advertising \cite{google2011arrival}. Unlike the traditional ad inventories buying by bundles, RTB enables the advertisers to perform the impression-level bidding based on the evaluation of the individual ad display opportunities. Such evaluation makes use of various real-time information such as the context data, user cookies and interest segments \cite{zhang2014optimal}. As a result, display advertising, especially its RTB part, brings the advertisers a higher return-on-investment (ROI), and thus more and more budget is allocated into display advertising \cite{IAB2013}.

Most display ads are charged based on \emph{pay per view} (PPV), i.e. each individual ad impression is charged \cite{evans2008economics,springborn2013impression}. The current measurement of ad impressions is simply implemented by the loading of the ad creatives. That is to say, when a webpage is loaded in the browser, each ad embedded in the page HTML is immediately counted as one impression. However, it is possible that some ad creatives are not viewed by the user, or even have not been exposed in the viewport (i.e. browser window in the screen). We refer such ad impressions as \emph{fraud impressions}. Because of the fraud impressions, publishers could earn some easy money by trying to add many ads at the bottom of the webpages. However, from a long-term perspective, the fraud impressions are definitely harmful for display advertising eco-system. Firstly, the advertisers have to pay for some ads never displayed, which seriously reduces the ROI. Secondly, the fraud impressions act as the noisy data in the machine learning training stage of impression evaluator in RTB platform. For example, a click-through rate (CTR) estimator is much important in RTB display advertising \cite{zhang2014optimal}. Its training data consists of the features of each ad impression with the label showing whether the user clicked the ad or not. If an ad impression is logged but the ad was actually not shown in the user viewport, such data is a noise which should reduce the prediction performance of the trained CTR estimator. With more wasted budget and inaccurate impression evaluation, the ROI of display advertising is reduced, which could possibly prevent the advertisers from allocating more budget into display advertising. As a result, it is much important to find an effective ad impression viewability measurement to get rid of such problems.

A perfect ad impression measurement should just match what the user has really viewed with at least a short memory. Firstly, the ad creative should be really shown in the viewport. Secondly, in order to have at least a short impression of the ad, the ad creative should be exposed in front of the user for a piece of time. Therefore, the two important factors we care about for measuring the ad impression viewability are: (i) \emph{pixel percentage}, i.e. the percentage of pixels of ad creative shown in the viewport, and (ii) \emph{exposure time}, i.e. how long the ad creative has been shown in the viewport.

In this paper, from a human-computer interaction perspective, we carefully study the relationship between the user's short memory on the ads with the two factors: pixel percentage and exposure time. We track the pixel percentage and exposure time of each ad creative using jQuery, based on which we develop 11 ad impression measurements. The user study experiments involve 20 participants, where each of them is asked to freely read some webpages with the tracked display ads. Then we check the match degree between the user's recalled ad impressions and the counted impressions from each measurement. Our empirical study shows that the pixel percentage and exposure time are both important factors for the user to remember the ads. And particularly, the exposure time significantly influences ad recall confidence. Specifically, the empirically best measurement counts an ad impression if 75\% pixels of the ad creative have been shown for at least 2 continuous seconds.

\section{Related Work}
Since the origin of online advertising, the researchers have been seeking a way to answer whether or not the user has viewed an ad in one webpage. In \cite{bayles2002designing}, the relationship between the ad creative animation and users' recognition were studied but the authors found such relationship was not significant. The authors in \cite{dreze2003internet,buscher2009you} leveraged the eye-tracking techniques to study the users' attention area on the screen and check how likely the user truly viewed a particular ad. Different users and webpages had much different high attention areas \cite{buscher2009you}. However, such eye-tracking techniques are impractical to be used in the production. The most practical and straightforward method is to define the pixel percentage and exposure time as the thresholds of measuring a viewed impression across all the webpages and users. In 2013, Google announced that the advertisers were charged only for the viewed ad impressions, where an ad was considered as viewed only if the pixel percentage was no less than 50\% and the exposure time was no less than 1 second\footnote{http://www.bbc.co.uk/news/business-25356956}. Later in 2014, IAB published a new standard for viewable display impressions, where the pixel percentage threshold was still 50\% while the exposure time threshold was 1 second for web display ads, and 2 seconds for video ads\footnote{http://www.iab.net/iablog/2014/03/viewability-has-arrived-what-you-need-to-know-to-see-through-this-sea-change.html}. It should be noted that such criteria is a result of balancing the benefit of advertisers, publishers and intermediaries. In our work, we study the ad impression viewability measurements purely from the human-computer interaction perspective.

\section{System Design}
The goal of our system is to investigate how the displayed pixel percentage and the exposure time influence the users' ad recall, and which impression viewability measurement best matches the users' remembered ad.

\subsection{Pixel Percentage Tracking}

\begin{figure}
\centering
\includegraphics[width=0.9\columnwidth]{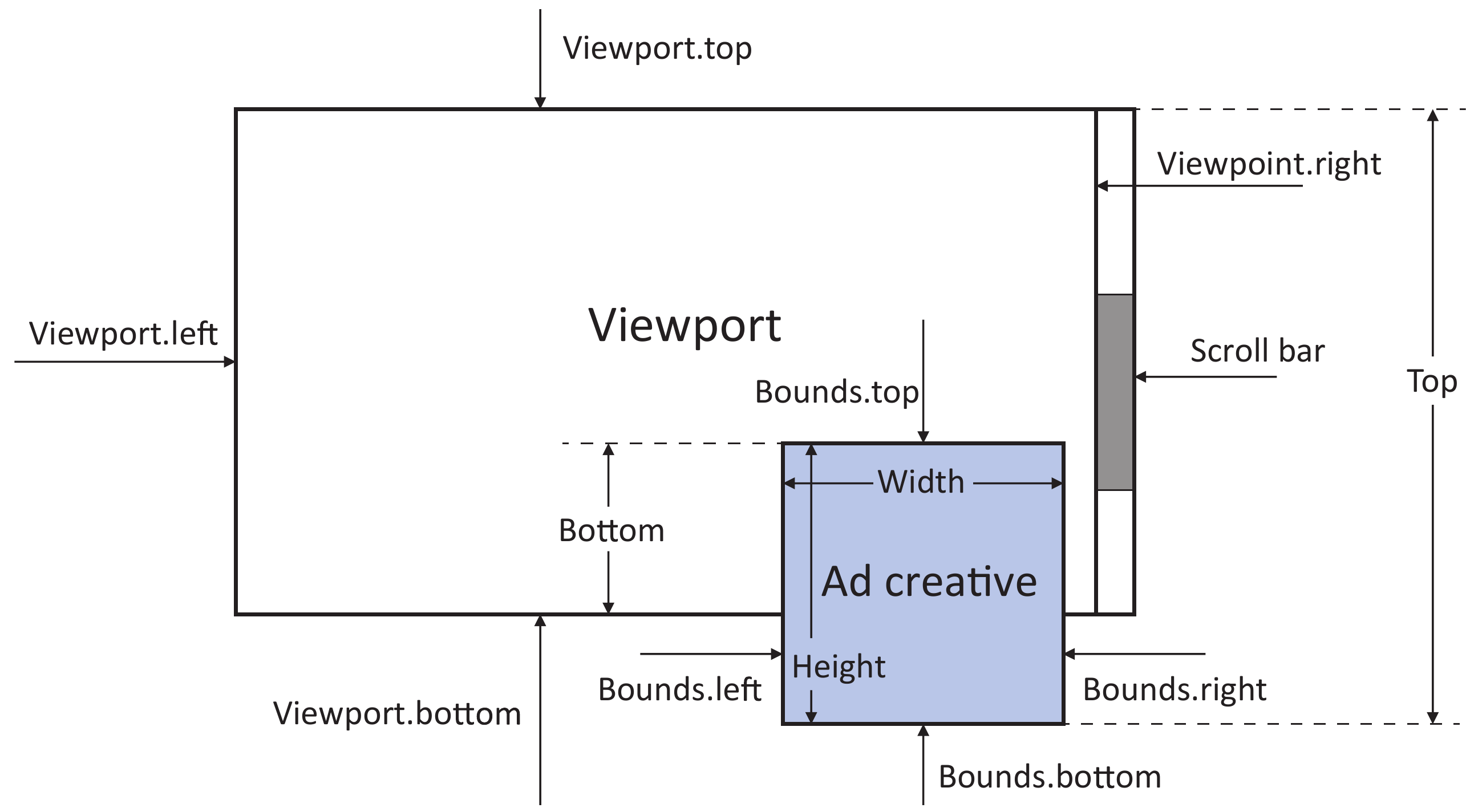}
\caption{Advanced pixel percentage tracking diagram \label{fig:adImp}}
\vspace{-10pt}
\end{figure}

The displayed pixel percentage for rectangle ad creative in the viewport can be calculated by the displayed height percentage times the displayed width percentage. Therefore, we tracked the bounds of browser's viewport and each ad creative.

Figure \ref{fig:adImp} shows the relationship of the variables. In the webpage coordinates, the upperleft point is the origin point. The lower place means the higher y-axis value and the right place means the higher x-axis value. Specifically, we calculate four ratios capped by 1:
{\scriptsize
\begin{align*}
\texttt{Top}& = \texttt{min(1, (bounds.bottom - viewport.top) / height)}\\
\texttt{Bottom}& = \texttt{min(1, (viewport.bottom - bounds.top) / height)}\\
\texttt{Left}& = \texttt{min(1, (bounds.right - viewport.left) / width)}\\
\texttt{Right}& = \texttt{min(1, (viewport.right - bounds.left) / width)}\\
\texttt{Pixel\%} & = \texttt{Top} \times \texttt{Bottom} \times \texttt{Left} \times \texttt{Right}
\end{align*}
}
In Figure~\ref{fig:adImp}, \texttt{Top}=\texttt{Left}=\texttt{Right}=1, \texttt{Bottom}=0.6, thus the pixel percentage is 60\%. Given an impression measurement with the pixel percentage threshold 50\%, the measurement will count this ad impression. Note that when any of the four factors is negative value, the entire ad creative is outside of the viewport, thus the pixel percentage is calculated as zero.

\subsection{Exposure Time Tracking}

The exposure time is associated with a pixel percentage threshold. For example, if the pixel percentage is 50\%, only after half pixels have been shown in the viewport does the tracking system start to count the exposure time. If we do not want any pixel percentage threshold, just set it as 0\%. If the measured exposure time has surpassed the predefined threshold, e.g., 2 seconds, then the measurement counts this ad impression.

Specifically, we used the tick counts based methods to calculate the exposure time.
For example, for the measurement of 50\% pixel percentage and 2 seconds exposure time, the tick counter will start to track the time once the pixel percentage meets 50\%. Then tick counter calls the pixel percentage tracking algorithm every 0.1 second for 20 times. Every time the pixel percentage tracking algorithm checks whether the current pixel percentage is no less than 50\%. If it returns false, the tick counter will restart the counting. If the tick counter counts up to 20, the exposure time and pixel percentage thresholds are both reached, thus the measurement counts this ad impression.

\section{Experiment Setting}
\subsection{Participants}
20 participants (10 male), students and staff at ANON-UNI, were recruited to take part in our user study. The average age was 23.2 (SD = 1.4). The majority spent 3 to 7 hours everyday using computer and had more than 8 year experience of using Internet.

\subsection{Materials}
\begin{figure}
\centering
\includegraphics[width=0.98\columnwidth]{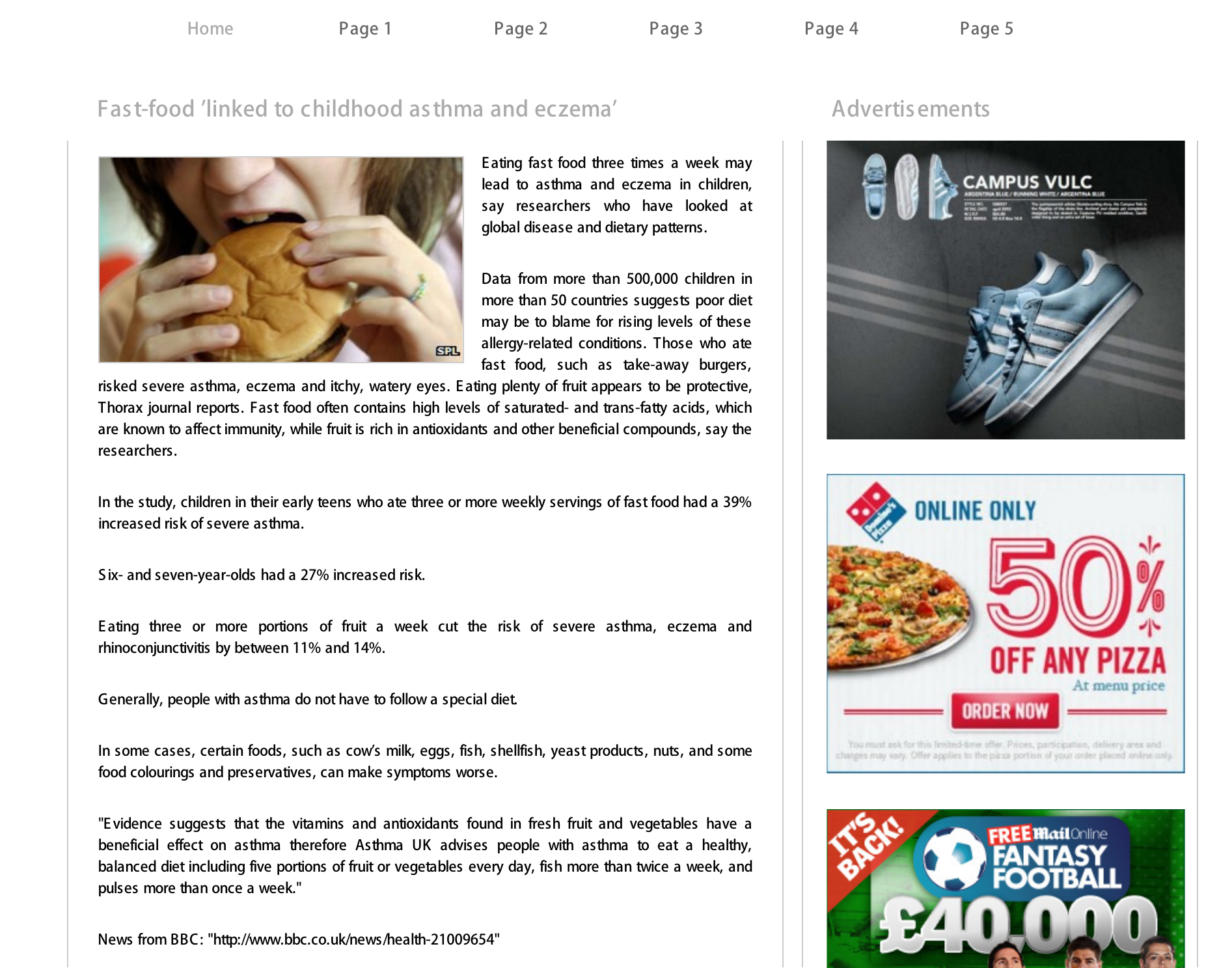}
\caption{Example test webpage.}
\label{fig:page}
\vspace{-15pt}
\end{figure}

We created 5 webpages with an article per page on the topic of sports, food, game, joke and gossip, respectively. We then randomly downloaded 60 ad creative images with uniform size of $300\times250$ pixels\footnote{According to IAB, this is one of the most standard display ad sizes. http://www.iab.net/wiki/index.php/Ad\_banner}. These ads covered various product categories, including electronic devices, food, sports, games, radio, shoes, education, gamble etc. We allocated 6 ads for each page, and deployed the impression tracking algorithm with all 11 measurements to each ad. In order to remove the position bias of each ad on the webpages, the ads were shuffled during page loading. The viewport on a sample webpage is shown in Figure \ref{fig:page}.

\subsection{Procedure}
Prior to starting the assessed part of the experiment, each participant completed the first questionnaire about their demographical information. For the next five minutes\footnote{We choose such time length according to the users' average stay time against the word number on webpages according to \cite{weinreich2008not}.}, each participant was asked to read the webpages freely without telling them the time limit for browsing the webpages and there would be questions about ads afterwards. After the participant completing browsing the webpages, they were presented with the second questionnaire. This questionnaire consisted of 60 ads with 30 appeared in the webpages and other 30 never shown. Participants were told to select the ads which they had just seen, and to give their subjective assessment for level of confidence in remembering each selected ad. The level of confidence was elicited on 7-point Likert scales with the anchor 1 (Not confident) - 7 (Strongly confident). Finally, they were presented with the post-experimental questionnaire relating to the level of understanding webpages' content as well as the degree of their attention to the ads while reading the webpages.

\subsection{Scoring}
The counting of each ad impression is a bi-classification problem. For each user-ad pair, the measurement judged whether the user had viewed this ad. And the labels were given by the user feedback on the second questionnaire.
We used precision, recall and F1 score to evaluate the impression counting performance of each measurement. Precision value is the fraction of counted relevant ad impressions over all the counted ones. Recall is the fraction of counted relevant ad impressions over all the user selected valid ad impressions. F1 score is calculated based on precision and recall: $F1=2PR/(P+R)$.

We also used the F1 score to evaluate the user's selection on valid impressions and article topics. The user might select some ads which were actually never occurred in the webpages. Also the user would select the wrong article topics.

\section{Empirical Results}
\subsection{Algorithm Performance}

\begin{figure}[t]
\centering
\subfigure{
\includegraphics[width=0.48\columnwidth]{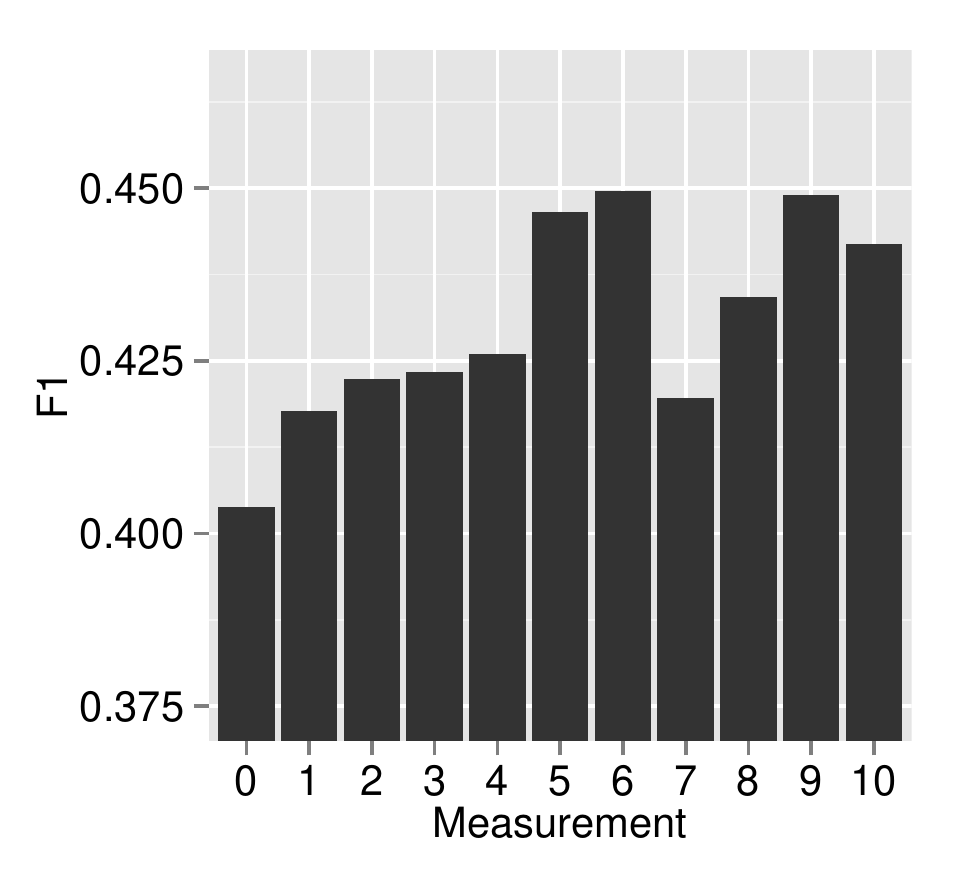}}
\subfigure{
\includegraphics[width=0.48\columnwidth]{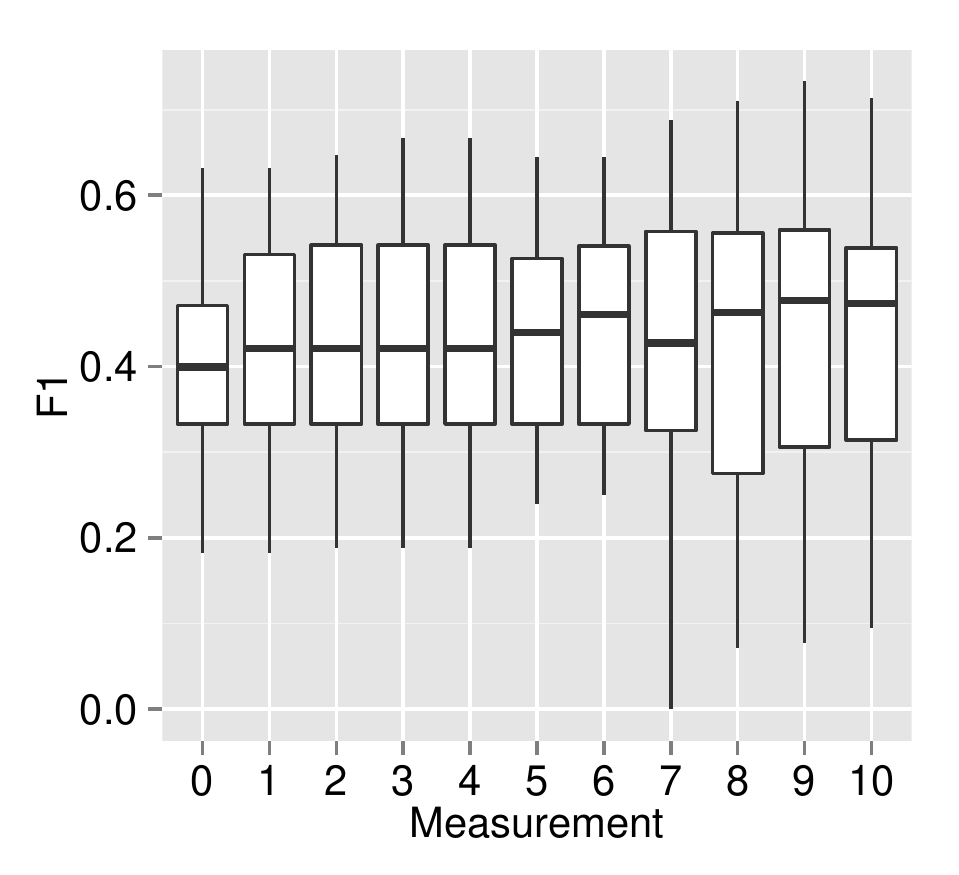}}
\small
\begin{tabular}{ccc|ccc|ccc}
meas. & pixel & time & meas. & pixel & time & meas. & pixel & time\\ \hline
0 & 0\% & 0 & 4 & 100\% & (0,2) & 8 & 50\% & [4,+)\\
1 & 0\% & (0,2) & 5 & 50\% & [2,4) & 9 & 75\% & [4,+)\\
2 & 50\% & (0,2) & 6 & 75\% & [2,4) & 10 & 100\% & [4,+)\\
3 & 75\% & (0,2) & 7 & 100\% & [2,4)\\
\end{tabular}

\caption{Average F1 and boxplot of F1 score for each measurement.}
\label{fig:stage-1-algo-f1}
%\vspace{-5pt}
\end{figure}

The F1 score and its distribution in box plot \cite{mcgill1978variations} for each measurement as well as the measurements' pixel percentage threshold and exposure time interval are presented in Figure~\ref{fig:stage-1-algo-f1}. We observed that measurement 6 ($\text{pixel} \geq 75\%, \text{time} \in [2,4)$) and measurement 9 ($\text{pixel} \geq 75\%, \text{time} \in [4,+\infty)$) provided the highest average F1 score and median F1 score. This suggested that an ad was probably viewed by the user if at least 75\% of its pixels had been shown in the viewport for no less than 2 seconds.

\begin{figure}[t]
\centering
\subfigure{
\includegraphics[width=0.9\columnwidth]{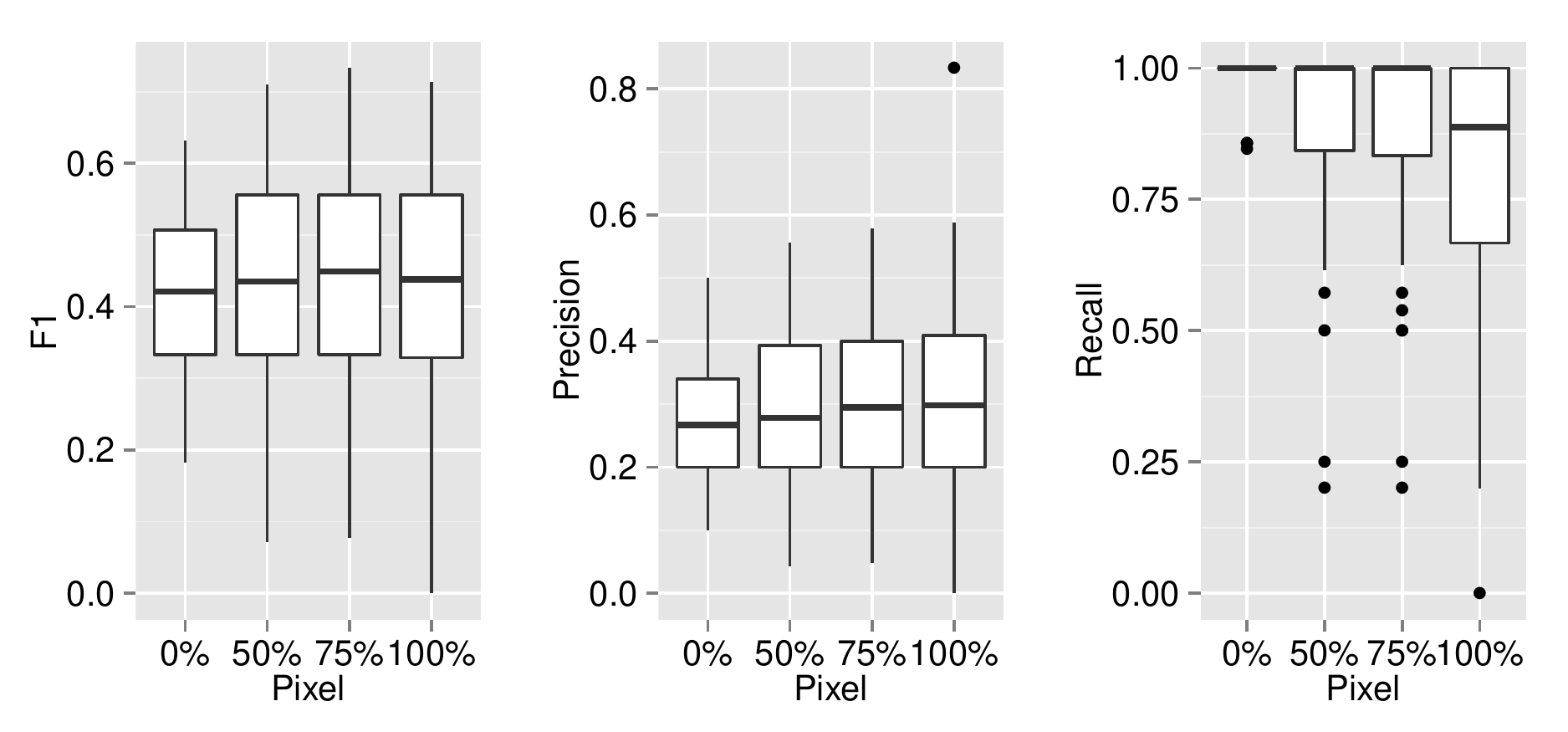}}
\subfigure{
\includegraphics[width=0.9\columnwidth]{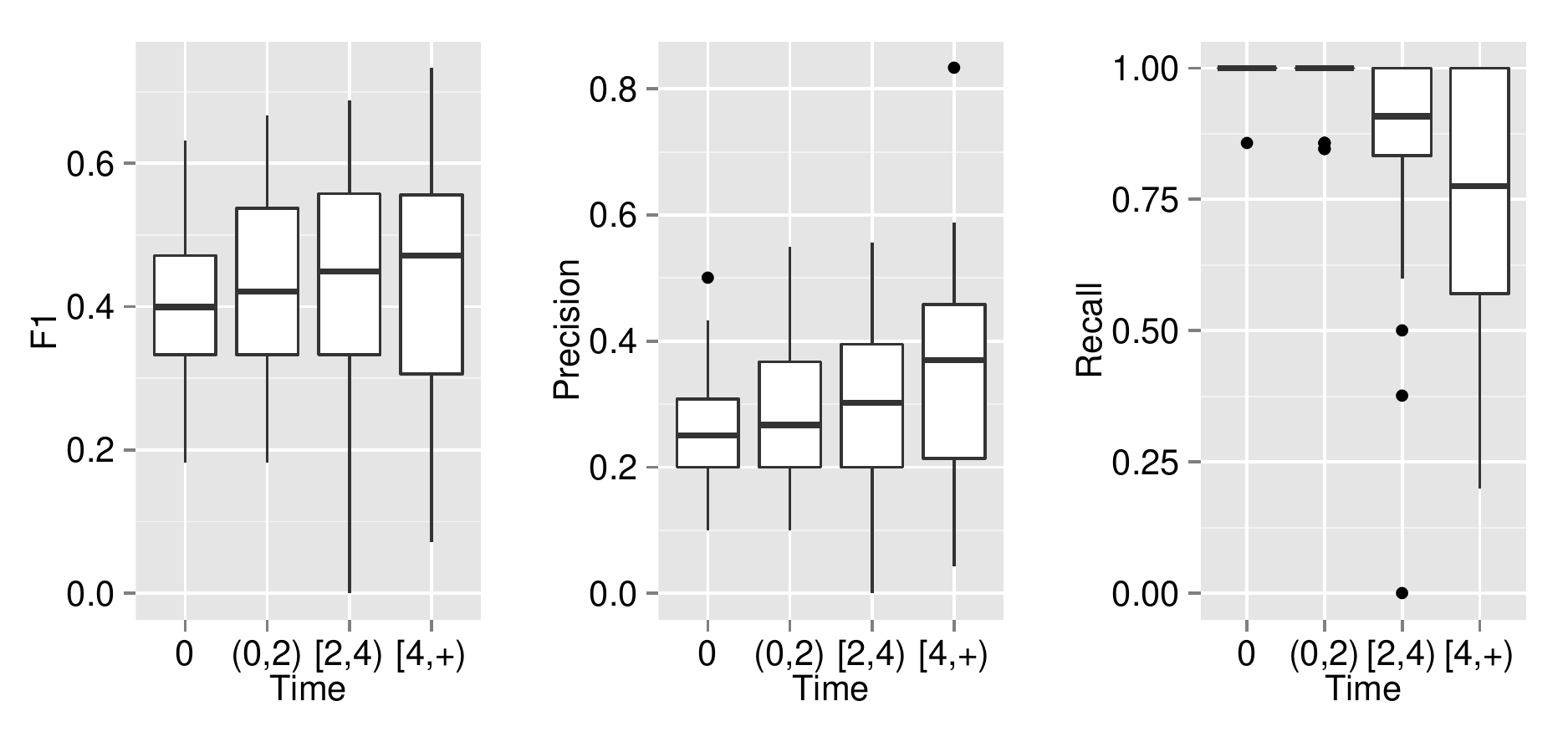}}
\caption{Boxplot of F1 score, precision and recall against each pixel percentage and exposure time.}
\label{fig:stage-1-pixel-f1-precision-recall-box-pixel-time}
\vspace{-5pt}
\end{figure}

In Figure~\ref{fig:stage-1-pixel-f1-precision-recall-box-pixel-time}, we further analysed the distribution difference of F1, precision and recall against different pixel percentages and exposure time respectively.
From the result we found that generally with the higher pixel percentage or the longer exposure time, the recall decreased whereas the precision increased. This finding is intuitive because the more ad display surface and the longer exposure time lead to the higher user ad recall. On the other hand, the higher pixel percentage and exposure time thresholds usually filtered out some of the truly viewed ads, leading to the lower recall. For the F1 score distribution, 75\% pixel percentage threshold and $[2,+\infty)$ exposure time leaded the highest F1, which was consistent with the individual measurement performance discussed above.

\subsection{User Analysis}

\begin{figure}[t]
\centering
\includegraphics[width=\columnwidth]{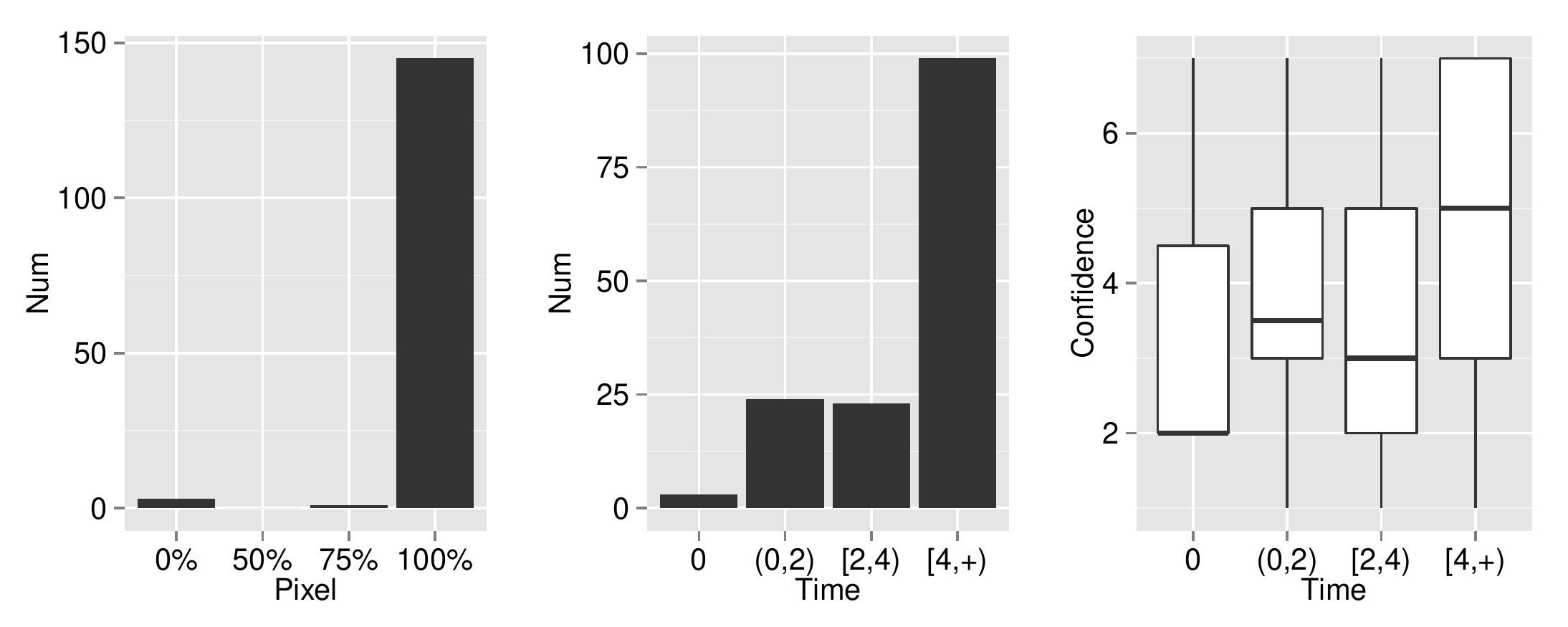}
\caption{Numbers of pixel percentage and exposure time for the user selected ads, with the user ad selection confidence on different exposure time.}
\label{fig:stage-2-user-conf}
\vspace{-10pt}
\end{figure}

We then investigated participants' self-reports of level of confidence in ad recall. Figure~\ref{fig:stage-2-user-conf} shows the numbers of each pixel percentage, exposure time interval and the participants' ad selection confidence against the exposure time.  We observed that 97.3\% of the user selected ad impressions were displayed with 100\% pixels, which meant the users almost just noticed on the ads that had been fully shown in the viewport\footnote{However, a quarter of the counted impressions with 100\% pixels only had less than 2 seconds exposure time, which corresponded to the measurement 4 with low performance. As a result shown in Figure~\ref{fig:stage-1-pixel-f1-precision-recall-box-pixel-time}, 75\% pixel overall worked better than 100\% pixel.}. On the exposure time dimension, most of the ad impressions were exposed for more than 4 seconds and in such case the user's ad selection confidence was much higher than that with shorter exposure time. A one-way ANOVA was conducted on confidence to compare the effect of exposure time. There was a significant effect of exposure time. $F(3, 145) = 4.042$, $p = .009$. Post hoc comparisons using the Tukey HSD test indicated that the mean score for the $\text{time} \in [4,+\infty)$ condition was significantly higher than the $\text{time} \in [2,4)$ condition, $p = .023$ . However, the $\text{time} \in [2,4)$ condition did not significantly differ from the $\text{time} \in [0,2)$ condition $p = .965$ , and the $\text{time} \in \{0\}$ condition $p = .999$.

\begin{figure}[t]
\centering
\includegraphics[width=\columnwidth]{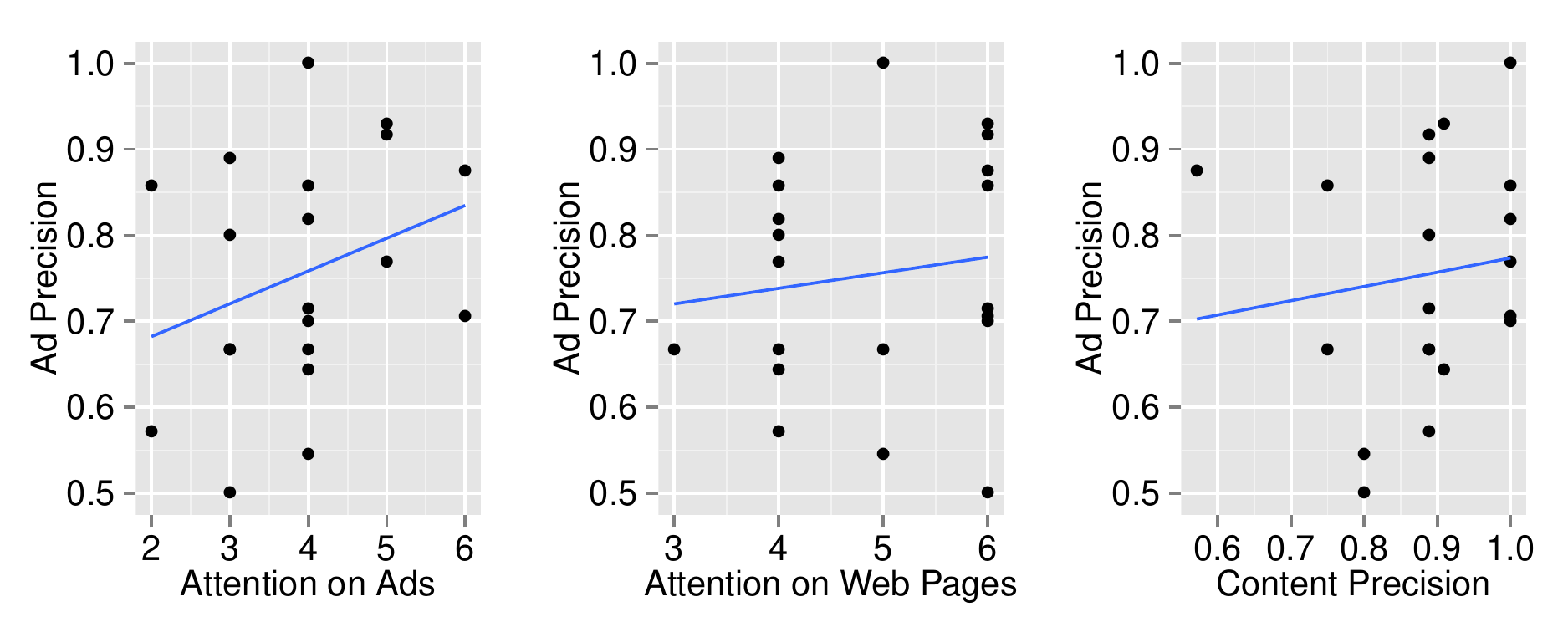}
\caption{Users ad selection precision against their attention on ads, content and content precision.}
\label{fig:stage-0-user-ad-precision}
\vspace{-5pt}
\end{figure}

It is possible that some user-selected ad impressions were in fact not shown in the webpages at all. In Figure~\ref{fig:stage-0-user-ad-precision}, we studied the precision of user ad selection against the users' reported levels of attention on ads, webpage content and their webpage topic selection precision. The relationships between ad selection precision and all these three variables were positive. Particularly, the last two subfigures in Figure~\ref{fig:stage-0-user-ad-precision} indicate an unintuitive finding that when the users pay more attentions on the webpage content, they will also unintentionally notice more ads.

\section{Conclusions}
In this paper, from the human-computer interaction perspective, we studied the contribution of ad impression viewability from the display pixel percentage and exposure time for an ad display opportunity. The results showed that with no less than 75\% of pixels being shown in the viewport for at least 2 continuous seconds, the ad impressions were effectively and sufficiently counted. Furthermore, comparing the dimensions of pixel percentage and exposure time, the study indicated that the exposure time acted as a more significant factor for the users' ad recall confidence on the ad impressions than the pixel percentage. In the future work, we plan to investigate more factors which could make difference on users' ad recall, such as the ad creative sizes, ad slot positions and ad image formats. And we plan to deploy our impression viewability measurement on a partner commercial display advertising platform to check whether the effectively ``viewed'' ad impressions would improve the advertising performance in a long term.

{
\bibliographystyle{acm-sigchi}
\bibliography{ad-tracking}
}
\end{document}